\documentclass[aps,12pt]{revtex4}

\begin{document}
\author{W. Zuo$^{1,2,3}$\footnote{Corresponding address:
Institute of Modern Physics, Chinese Academy of Science, P.O.Box
31, Lanzhou 730000, China. Tel: 0086-931-4969318; E-mail:
zuowei@impcas.ac.cn}, Z.H. Li$^{1,2}$, A.Li$^3$, U. Lombardo$^{4}$}
\affiliation{$^1$ Institute of Modern Physics, Chinese Academy of Sciences,
Lanzhou 730000, P.R. China\\
$^2$ Graduate School of Chinese Academy of Sciences,
Beijing 100039, P. R. China\\
$^2$ School of Physics and Technology, Lanzhou University,
Lanzhou 730000, P.~R.~China\\
$^4$ INFN-LNS, Via Santa Sofia 44, I-95123 Catania, Italy}
\title{Effect of Three-body Interaction on Phase Transition of
Hot Asymmetric Nuclear Matter}

\begin{abstract}
The properties and the isospin dependence of the liquid-gas phase transition
in hot asymmetric nuclear matter have been
investigated within the framework of the finite temperature
Brueckner-Hartree-Fock approach extended to include the
contribution of a microscopic three-body force. A typical Van der Waals
structure has been observed in the calculated isotherms (of pressure) for
symmetric nuclear matter implying the presence of the liquid-gas phase
transition. The critical temperature of the phase transition is calculated
and its dependence on the proton-to-neutron ratio is discussed.
It is shown that the three-body force gives a repulsive contribution
to the nuclear equation of state and reduces appreciably the critical temperature
and the mechanical instable region.
At fixed temperature and density the pressure of asymmetric nuclear matter increases
monotonically as a function of isospin asymmetry.
In addition, it turns out that the domain of mechanical instability for
hot asymmetric nuclear
matter gradually shrinks with increasing asymmetry and temperature.
We have compared our results with the predictions of other theoretical
models especially the Dirac
Brueckner approach. A possible explanation for
the discrepancy between the values of the critical temperature predicted
by the present non-relativistic Brueckner calculations including the three-body
force and the relativistic
Dirac-Brueckner method is given.
\\[2mm]
\end{abstract}
\maketitle

\section{Introduction}

The determination of the properties of nuclear matter as a function of
temperature, density and isospin asymmetry
is one of the fundamental subjects in nuclear physics.
For a long time it has been expected that, due to the Van der Waals
nature of the nucleon-nucleon (NN) interaction, nuclear matter is likely to
exhibit a liquid-gas phase transition at low densities and moderate
temperatures\cite{bertsch:1983}. Phase transition is a phenomenon of
great and extensive interests observed or expected in various fields,
such as melting or boiling in condensed matters
and atomic clusters\cite{schmidt:2001,gobet:2001},
decay of hot nuclei\cite{lopez:2001,gupta:2001} and
transition to quark-gluon plasma in high energy heavy
ion collisions and neutron stars\cite{matsui:1986}.
 In heavy ion collisions at
intermediate and high energies, hot asymmetric nuclear matter can be
produced and during the expansion stage
of the collisions the liquid-gas phase transition may occur
depending on the temperatures and densities
involved\cite{bonche:1984,muller:1995}.
Besides its general interest in nuclear physics, the equation of
state (EOS) of hot asymmetric nuclear matter
plays also an essential role in understanding many astrophysical phenomena
related to the dynamics of supernova explosions and the evolution of
the protoneutron star formed in the latest stage of a type-II supernova
collapse\cite{prakash:1997}.
It is therefore highly desirable to predict on a microscopic basis
the thermal properties of asymmetric nuclear matter at finite temperature
which provide a starting point for the physics of heavy
ions\cite{stock:1982,fuchs:1997}
and protoneutron stars\cite{baron:1985}.

Within various theoretical models such as the phenomenological
non-relativistic Skyrme force
models\cite{jaqaman:1983,lattimer:1985}
 and the relativistic
mean field theory (RMT)\cite{muller:1995}, the EOS of hot
nuclear matter has been studied extensively.
All of these investigations have predicted a Van der Waals behavior for
infinite symmetric nuclear matter.
The critical temperature $T_c$
of the liquid-gas phase transition turns out to be
in the range from 14 MeV to 20 MeV depending on
the adopted models and NN interactions.
Based on realistic NN interactions, the EOS of cold nuclear matter has
been explored by many physicists within the variational
method\cite{wiringa:1988},
the non-relativistic
Brueckner-Hartree-Fock (BHF) approach\cite{bombaci:1991,zuo:1999} and the
Dirac Brueckner (DB) theory\cite{alonso:2003,fuchs:2004}.
However, the investigations of hot asymmetric nuclear matter
from the microscopic models are relatively rare.
In Ref.\cite{friedman:1981}, the liquid-gas phase
 transition has been predicted for
symmetric nuclear matter by using
the variational method based on the
Argonne $V_{14}$ two-body NN interaction plus a phenomenological three-body
force and the critical temperature obtained is about 17.5 MeV close to the values
from the Skyrme force calculations.
On the contrary, it is reported in Ref.\cite{haar:1987} from
 the finite temperature
DB calculations an almost complete destruction of Van der Waals like
behavior. As been discussed in Ref.\cite{baldo:1995}, this unusual result
in the DB framework has not been well understood and deserves deeper
investigations. In a more recent work\cite{huber:1998}, the Van der Waals
structure has been obtained in the DB framework, but the predicted critical
temperature is as low as $\sim 10 MeV$.
In Ref.\cite{bombaci:1994},
 the properties of hot asymmetric nuclear matter
has been studied in the finite temperature BHF approach by adopting
the separable version of the Paris two-body force.
As well known, within a non-relativistic microscopic framework
three-body forces are decisive for reproducing the empirical
saturation properties of cold nuclear matter and in the zero-temperature case
their contribution shifts the EOS close to the one of the relativistic DB
calculations\cite{zuo:2002}.
In the present paper, our aim is to explore the EOS of hot
asymmetric nuclear matter in the finite temperature
BHF approach extended to include the
contribution of the microscopic three-body force (TBF) constructed from the
meson exchange current model\cite{grange:1989}.
We shall stress the isospin dependence of the liquid-gas phase transition and
the effect of the TBF in particular its possible connection to the
relativistic effect in the
DB approach. In the present calculations, the Argonne $V_{18}$ ($AV_{18}$)
potential\cite{wiringa:1995} has been
adopted as the realistic two-body interaction.
The paper is arranged as follows. In the next
section we give a brief description of the theoretical approaches.
The numerical results are presented and discussed in Section III.
Finally a summary is given in Sect.IV.

\section{ Theoretical Models }
\subsection{Temperature-dependent Effective Three-body Force}

The microscopic TBF adopted in the present calculations is
constructed from the meson-exchange current approach~\cite{grange:1989}
and its components are displayed diagrammatically in Fig.\ref{fig1}, taken
from Ref.\cite{grange:1989}. Four important mesons $\pi$, $\rho$, $
\sigma $ and $\omega $ are considered\cite{machleidt:1989}. The TBF model
contains the contribution of the two-meson exchange part of the NN
interaction medium-modified by the intermediate virtual excitation
of nucleon resonances, the term associated to the non-linear
meson-nucleon coupling required by the chiral symmetry, the
simplest contribution rising from meson-meson interaction and
finally, the two-meson exchange diagram with the virtual
excitations of nucleon-antinucleon pairs. The meson masses in the
TBF have been fixed at their physical values except for the
virtual $\sigma$-meson mass which has been fixed at $540$MeV
according to Ref.~\cite{grange:1989}. This value has been checked to
satisfactorily reproduce the $AV_{18}$ interaction from the
one-boson-exchange potential (OBEP) model\cite{zuo:2002}. The other
parameters of the TBF, i.e., the coupling constants and the
form factors, have been determined from the OBEP
model to meet the self-consistent requirement with the adopted
$AV_{18}$ two-body force. The parameters of the TBF is
given in Ref.\cite{zuo:2002}. A more detailed description of the
model and the approximations can be found in
Refs.~\cite{grange:1989,zuo:2002}.

In the zero-temperature case, the TBF contribution has been included in the
BHF calculations by constructing an effective two-body interaction via
a suitable average with respect to the third-nucleon degrees of
freedom\cite{grange:1989,lejeune:1986}.
By extending the standard scheme adopted for the zero-temperature case,
one can reduce the TBF to a temperature dependent effective two-body
force $V_3^{\rm eff}(T)$ which reads in $r$-space,
\begin{eqnarray}\label{e:tbf}
 \langle \vec r_1^{\ \prime} \vec r_2^{\ \prime}| V_3^{\rm eff} (T)|
\vec r_1 \vec r_2 \rangle &=& \displaystyle
 \frac{1}{4} {\rm Tr} \sum_{k_n} f(k_n,\rho,\beta,T) \int {\rm d}
{\vec r_3} {\rm d} {\vec r_3^{\ \prime}}\phi^*_n(\vec r_3^{\
\prime}) (1-\eta(r_{13}', T ))\nonumber
 \\[2mm] & \times &
\displaystyle (1-\eta(r_{23}', T))
W_3(\vec r_1^{\ \prime}\vec r_2^{\ \prime} \vec r_3^{\ \prime}|\vec
r_1 \vec r_2 \vec r_3) \\ \nonumber
&\times &
\phi_n(r_3) (1-\eta(r_{13}, T))
 (1-\eta(r_{23}, T)) ,
\end{eqnarray}
where the trace is taken with respect to spin and isospin of the third
nucleon. The function $\eta (r,T)$ is the defect
function\cite{grange:1989,lejeune:1986,baldo:1999}
which is defined as $\eta (r,T) = \phi (r) - \psi (r,T)$, where $\psi (r,T)$
is the correlated wave function for the relative motion of two nucleons in
nuclear medium and $\phi(r)$ is the corresponding unperturbed one.
A detailed description and justification of the above scheme
can be found in Ref.\cite{grange:1989}.

It is worth stressing that the TBF itself, i.e. $W_3$, is the same as the one
adopted in our previous calculations for the zero-temperature
case\cite{zuo:2002} and it is independent of temperature.
However, in the finite-temperature case, the
effective two-body force $V_3^{\rm eff}(T)$ constructed from the
TBF depends on temperature in an implicitly complicated way
due to the medium effects.
As a consequence the contribution of the TBF is expected to be more
pronounced at finite temperature.
It is clear from Eq.(\ref{e:tbf}) that the
temperature dependence of $V_3^{\rm eff}(T)$ stems from the Fermi
distribution $f(k,\rho,\beta,T)$ and the defect function
$\eta(r,T)$ which is strongly temperature dependent. We will return to
this point in the next subsection.
\subsection{Finite Temperature Brueckner-Hartree-Fock Approach}
In general, three independent parameters are required to specify a given
thermodynamical state of hot asymmetric nuclear matter, i.e., the
total nucleon number density $\rho$, the isospin asymmetry
parameter $\beta =(\rho_n-\rho_p)/\rho$ and the temperature
$T$.
At zero temperature, the neutron and proton Fermi momenta
are related to their respective number densities
$\rho_n=(1+\beta )\rho/2$ and
$\rho_p=(1-\beta )\rho/2$ by
$k_F^\tau =[3\pi ^2\rho _\tau ]^{1/3}$ with $\tau =p$ or $\tau =n$.
The Brueckner-Bethe-Goldstone (BBG) approach for cold asymmetric nuclear
matter is described in Ref.\cite{bombaci:1991,zuo:1999}.
The extension to finite temperature is given in Ref.\cite{bombaci:1994}.
In the following, we
give a brief review for completeness.
 The starting point of the BBG
approach is the Brueckner reaction $G$ matrix which satisfies the
following generalized Bethe-Goldstone (BG) equation\cite{bombaci:1991},
\begin{equation}\label{e:ftbg}
G_{\tau ,\tau ^{\prime }}(\rho ,\beta, T, \omega )=v+v\sum\limits_{k_1k_2}%
\frac{\mid k_1k_2\rangle Q_{\tau ,\tau ^{\prime }}(k_1,k_2)\langle k_1k_2\mid }{%
\omega -\epsilon_{\tau}(k_1)-\epsilon_{\tau'}(k_2)}G_{\tau ,\tau ^{\prime
}}(\rho, \beta, T,\omega ) ,
\end{equation}
where $v=v_2+V_3^{\rm eff}(T)$ is the NN
interaction and $\omega $ is the starting energy. In the present
calculations the $AV_{18}$ potential is adopted as the bare two-body
force $v_2$ and the TBF contribution is included via the effective
interaction $V_3^{\rm eff}(T)$ given by Eq.(\ref{e:tbf}).
Since the defect function is determined by the $G$ matrix,
the effective force $V_3^{\rm eff}$ should be evaluated self-consistently
with the BG equation at each step of the Brueckner iteration.
In Eq.(\ref{e:ftbg}), the single particle (s.p.) energy is defined as
$\epsilon_\tau (k)\equiv \epsilon_\tau (k,\rho ,\beta,T)=\hbar ^2k^2/(2m)+
U_\tau (k, \rho, \beta, T)$.
In the present calculations, we adopt
the continuous choice\cite{jeukenne:1976} for
the s.p. potential $U_\tau (k)\equiv U_\tau (k,\rho,\beta, T)$ since it
is an natural choice for $T\ne 0$\cite{baldo:1999,baldo:1988} and
it provides a much faster convergence of the hole-line expansion
in the zero-temperature limit than the gap choice\cite{song:1998}. The s.p.
potential is calculated from the real part of on-shell $G$ matrix,
\begin{eqnarray}\label{e:spp}
U_\tau (k,\rho, \beta, T) = \frac{1}{2} \sum_{\tau^{\prime}}
\sum_{\vec k^{\prime}} f_{\tau^{\prime}}(k^{\prime},\rho,\beta, T)
\langle kk^{\prime} \mid G_{\tau,\tau^{\prime }}(\rho, \beta, T,
\epsilon_{\tau}(k)+\epsilon_{\tau^{\prime}}(k^{\prime}))\mid
kk^{\prime}\rangle_A ,
\end{eqnarray}
where the subscript $A$ denotes the anti-symmetrization of the matrix
elements. The finite temperature Pauli operator is simply an
extension of the zero-temperature one, i.e.,
\begin{equation}\label{e:fq}
Q_{\tau ,\tau ^{\prime }}(k_1,k_2)\equiv
Q_{\tau ,\tau ^{\prime }}(k_1, k_2,\rho,
\beta,T)=[1-f_\tau (k_1,\rho ,\beta,T )][1-f_{\tau ^{\prime
}}(k_2,\rho ,\beta, T)].
\end{equation}
In Eq.(\ref{e:ftbg}), the Pauli operator is applied only for the
intermediate
states of momenta $k_1, k_2$.
The Fermi distribution for $T\ne0$ is expressed as,
\begin{equation}
f_\tau (k,\rho,\beta,T)=\left[1+\exp\left(\frac{\epsilon_\tau(k)
-\mu_\tau }T\right)\right]^{-1} ,
\end{equation}
where $\mu_\tau =\mu_\tau (k,\rho,T)$ is the chemical
potential. For any given
density and temperature, we can calculate the chemical potential
$\mu_\tau $ from the following implicit equation
self-consistently by iteration,
\begin{equation}\label{e:den}
\rho_\tau =\frac 1V\sum_kf_\tau (k,\rho,\beta,T)=\frac 1V\sum_k
\left[1+\exp\left(\frac{e_\tau(k)-\mu_\tau }T\right)\right]^{-1}.
\end{equation}
The BG equation can be expended in
partial waves as in the zero-temperature case\cite{baldo:1999}.
The resulting equations in different partial waves are coupled
with each other due to the angular dependence of the Pauli operator
and the s.p. energy.
To remove the angular dependence, we approximate the exact
Pauli operator and s.p. energy in the BG equation by their
angle-averaged values. For instance, the angle average of the
Pauli operator is expressed as,
\begin{equation}\label{e:qav}
\langle Q_{\tau \tau ^{\prime }}(q,P,\rho,\beta,T)\rangle
=\frac 12\int_0^\pi
\sin \theta d\theta [1-f_\tau (k_1,\rho,\beta,T)][1-f_{\tau ^{\prime
}}(k_2,\rho,\beta,T)] ,
\end{equation}
where $\vec q =(\vec k_1 - \vec k_2)/2$ and
$\vec P = \vec k_1 + \vec k_2 $ are the relative momentum and the total
momentum of the two nucleons, respectively. $\theta$ is the angle between
$\vec{q}$ and $\vec{P}$.
In the case of $T=0$, the angle-average of the Pauli operator
 can be derived analytically due to the sharp Fermi distribution
 at $T=0$\cite{bombaci:1991,zuo:1999,alonso:2003}. Whereas
 for $T\ne 0$, the integral in Eq.(\ref{e:qav})
 can not be worked out explicitly and one has to evaluate the
 angle-average of the Pauli operator numerically.
 In the present paper, the integral in Eq.(\ref{e:qav}) is calculated
 numerically by the Gauss-Legendre method.
 We have checked that in the zero-temperature
 limit our numerical results can reproduce the analytical values
 with very high accuracy.

The energy per nucleon of asymmetric nuclear matter at the BHF level of
approximation is given by,
\begin{eqnarray}\label{e:energy}
E(\rho,\beta,T)&=&\sum_\tau \sum_kf_\tau
(k,\rho,\beta,T)\frac{\hbar ^2k^2}{2m}
+ \frac 12\sum_{\tau,\tau ^{\prime
}}\sum_{k,k^{\prime}}f_\tau (k,\rho,\beta,T)f_{\tau ^{\prime
}}(k^{\prime},\rho,\beta,T)
 \nonumber \\
&\times &
 \langle kk^{\prime}\mid G_{\tau,\tau
^{\prime }}(\rho,\beta,T,\epsilon(k)+\epsilon(k^{\prime}))\mid kk^{\prime
}\rangle_A .
\end{eqnarray}
At the mean field approximation, the total entropy $S$ is expressed
as\cite{bombaci:1994}
\begin{equation}\label{e:entropy}
S=-\sum_{\tau}\sum_{k}\{f_{\tau}(k,\rho ,\beta ,T)
\ln f_{\tau}(k,\rho, \beta, T)+
[1-f_{\tau}(k,\rho,\beta ,T)]\ln [1-f_{\tau}(k,\rho ,\beta ,T)]\}
\end{equation}
The free energy $F$ is calculated according to the
standard thermodynamic relation $F=E-TS$ and the pressure
$P$ is then extracted from the free energy by performing a
numerical derivative, i.e.,
\begin{equation}
P =\rho ^2\left(\frac{\partial F}{\partial \rho }\right)_{T,\beta } .
\end{equation}
According to Ref.\cite{muller:1995} the criteria for mechanical
stability can be straightforwardly expressed as follows
$$
\rho ^2\left(\frac{\partial ^2F}{\partial \rho ^2}\right)_{T,\beta }
=\rho \left(\frac{\partial P}{\partial \rho }\right)_{T,\beta }>0 .
$$
The set of coupled equations (\ref{e:ftbg}), (\ref{e:spp}) and (\ref{e:den})
is referred to as the finite temperature BHF
approximation (FTBHF)\cite{bombaci:1994}.
When the TBF contribution is included, this set of equations has to be solved
self-consistently along with Eq.(\ref{e:tbf}) to get the $G$ matrix.
Hereafter we will call the FTBHF including the TBF contribution as the
extended FTBHF.

As a check of the extended model, we
have calculated the speed of sound in cold symmetric nuclear matter, i.e.,
$s/c=\left(\frac{dP}{d\varepsilon }\right)^{1/2}$, where $c$
represents the speed of light and $\varepsilon $ refers to the total
 energy density. The results are shown in Tab.~1 for
several densities. In the table
the result including the TBF contribution
is denoted by ${\rm BHF} (AV_{18} +$ TBF), while the one without the TBF
by ${\rm BHF} (AV_{18}$).
It is seen that in both cases with and without the TBF,
the calculated speed of sound fulfills
the causality condition $s/c<1$ in the relevant density region.
Below a certain value of
density ($\sim 0.14$fm$^{-3}$ in the case without the TBF and
$\sim 0.11$ fm$^{-3}$ with the TBF) the speed of sound $s/c$
becomes imaginary due to
the mechanical instability.
\begin{table}[ht]
\begin{center}
\caption{Speed of sound $s/c$ in symmetric nuclear matter for
several values of density in both cases, with and without the
TBF. }
\end{center}
\begin{center}
\begin{tabular}{|c|c|c|c|c|c|c|c|}
\hline
   $\rho ({\rm fm}^{-3})$
   & ~ 0.15 ~ &~ 0.20 ~&~ 0.25 ~&~ 0.30 ~&~ 0.35 ~&~ 0.40 ~&~ 0.45 ~\\
\hline
${\rm BHF} (AV_{18})$ & 0.01 & 0.08 & 0.15 & 0.20 & 0.25 & 0.29 & 0.35 \\
${\rm BHF} (AV_{18} +$ TBF) & 0.07 & 0.16 & 0.24 & 0.33 & 0.42 & 0.51 & 0.59 \\
\hline
\end{tabular}
\end{center}
\end{table}

\section{ Numerical Results and Discussions}
The EOS of asymmetric nuclear matter is reported in Fig.\ref{fig2}
for four values of asymmetry parameter $\beta=0, 0.3, 0.5, 0.8$,
where the solid and dashed isothermal curves of pressure ( corresponding to
$T=0,8,10,12,14,16MeV$ from the bottom to the top) indicate the
results with and without the TBF contribution, respectively.
From the figure one can see that for both cases with and without
the TBF, the EOS of symmetric nuclear matter ($\beta=0$) displays a typical
Van der Waals behavior, implying that the infinite
nuclear system may undergo a liquid-gas phase transition.
The critical temperature and density of the phase transition is
determined by the condition
$\left(\frac{\partial P}{\partial \rho }\right)_{T,\beta}=
\left(\frac{\partial^2P}{\partial \rho ^2}\right)_{T,\beta}=0$.
From the calculated isotherms of pressure, one can extract the critical
temperature $T_c$ and the critical density $\rho_c$. In the case without
the TBF contribution, the obtained $T_c$ is approximately $16$ MeV
which is in the range of $15-20$ MeV predicted by
the Skyrme-Hartree-Fock (SHF) calculations\cite{jaqaman:1983}.
As expected, the TBF gives a repulsive contribution to the
nuclear EOS. In the zero-temperature case, the additional repulsion from
the TBF improves greatly the predicted saturation density of cold
symmetric nuclear matter. The equilibrium properties of cold nuclear
matter have been reported in Ref.\cite{zuo:2002}.
At higher temperature, the TBF effect becomes
more pronounced due to the temperature dependence of its contribution
(i.e., the effective force $V_3^{\rm eff}(T)$). Inclusion of the TBF
contribution reduces the critical temperature from $\sim 16$ MeV to
$\sim 13$ MeV which is close to the value of $\sim 14$ MeV from the
relativistic mean field theory\cite{muller:1995}.
The reduction of $T_c$ due to the TBF may be attributed largely
to the medium and temperature dependence of the reduced effective interaction
$V_3^{\rm eff}(T)$, i.e., at a fixed temperature the TBF contribution is
stronger for larger density and for a given density it is more pronounced
at higher temperature.
The calculated critical density $\rho_c$ of the phase transition is roughly
$0.065$ fm$^{-3}$ and $0.08$ fm$^{-3}$ in the two cases with and without
the TBF, respectively. Both values are in the density range of
$\rho_0/3 \sim \rho_0/2$ in agreement with
other investigations\cite{muller:1995,jaqaman:1983,lattimer:1985,huber:1998}.
The TBF reduces the critical density since its
contribution to the nuclear EOS is repulsive and increases as a
function of density.

Now let us compare our result with the predictions of other approaches.
In Ref.\cite{friedman:1981}, the liquid-gas phase
 transition of hot nuclear matter
has been studied with the variational method by adopting the Argonne $V_{14}$
two-body NN interaction supplemented with a phenomenological TBF.
The obtained value of $T_c$ in Ref.\cite{friedman:1981}
 is about 17.5 MeV which
is higher than the present value $T_c\sim 13$MeV from the BHF approach
when the TBF contribution
is included. This disagreement concerning the critical temperature
might be attributed to the two different approaches and to the
different NN interactions adopted.
 As it has been noticed in Ref.\cite{baldo:1997}, the
density dependence of the symmetry energy predicted from the two methods
differs remarkably from each other even when the same NN
interaction has been used. The reason for the discrepancy between the
 results of the
two methods remains largely unclear as discussed in Ref.\cite{zuo:1999}.

In the relativistic DB framework, the EOS of hot
nuclear matter has been investigated in Refs.\cite{haar:1987,huber:1998}
where the extension to finite temperature has been achieved by the
inclusion of the finite temperature Green's functions\cite{serot:1986}.
Such an extension coincides in the non-relativistic limit with the one
adopted in the present FTBHF calculations. In Ref.\cite{haar:1987}, the
reported
EOS does not display any Van der Waals like behavior in contrast to the
results obtained from other approaches such as the non-relativistic BHF,
the SHF, the variational method, and the RMT.
As discussed in Ref.\cite{baldo:1995} this discrepancy is not easy to be understood
since the inclusion of the TBF in the BHF calculations is expected to
give a similar behavior of the EOS to that of the DB approach.
One possible reason for that is that in the calculations of Ref.\cite{haar:1987}, the
$\Delta$-resonance has been included explicitly which produces an additional
repulsion to the EOS and makes nuclear matter less bound.
In a later work\cite{huber:1998},
the problem has been re-visited in the DB approach
and a clear Van der Waals structure is observed in the calculated EOS.
However, the obtained critical temperature $T_c\simeq 10$ MeV
in Ref.\cite{huber:1998} is considerably lower
 than the value of $\sim 13$ MeV in the
present work.
As pointed out in Refs.\cite{baldo:1995,brown:1987} that the main
 relativistic effect in the
DB approach is associated to the contribution (of the TBF) due to the
$2\sigma$-exchange process coupled to the virtual excitation
 of a nucleon-antinucleon
pair which is referred to as $2\sigma$-$N\overline{N}$ TBF in
Refs.\cite{zuo:2002,grange:1989}.
In our previous investigations\cite{zuo:2002},
it is shown that the most important relativistic
correction to the nuclear EOS can be fairly well reproduced by
the BHF calculation including only the $2\sigma$-$N\overline{N}$
 contribution of the TBF.
This implies that the full TBF in the present calculations is not
completely equivalent to the relativistic effect in DB approach especially
in the finite-temperature case for which both the relativistic and the TBF effects
are expected to become more pronounced as compared to the zero-temperature case.
In order to get a deeper insight into the
disagreement of the predicted critical temperature between the DB approach
and the FTBHF approach with the TBF contribution, we
separate the $2\sigma$-$N\overline{N}$ contribution from the full TBF.
The results are given in Fig.\ref{fig3} where the solid curves
are obtained by including only the $2\sigma$-$N\overline{N}$ contribution
and the dashed curves by adopting the full TBF. It is seen that
the contribution of the full TBF is somewhat less repulsive than that of
the $2\sigma$-$N\overline{N}$ component. As a consequence, the value of $T_c$
is lowered from $\sim 13$ MeV to $\sim 11$ MeV if only
 the $2\sigma$-$N\overline{N}$
contribution in the TBF is taken into account.  This result is compatible with
our previous one for the zero-temperature case\cite{zuo:2002} and
 provides a possible explanation for the
above mentioned discrepancy.

As well known, pure neutron matter is unbound, therefore it is interesting
to discuss the isospin dependence of the phase transition.
By comparing the results for different isospin
asymmetries (Fig.\ref{fig2}a,b,c and d),
one can see that as the isospin asymmetry increases, the Van
der Waals structure of the EOS becomes less pronounced and the
mechanical instable region where the pressure decreases as a
function of density, gets smaller.
This indicates that the critical temperature drops down monotonically as the asymmetry
increases in both cases with and without the TBF, in agreement with
the results obtained from the isospin lattice gas model\cite{ray:1997}.
The TBF
gives a repulsive contribution to the EOS. The repulsion of the TBF becomes stronger
as the density increases and consequently reduces the critical temperature in
the whole range of asymmetry.
When the asymmetry is high enough, the region of mechanical instability vanishes
for all values of temperature. The extracted critical asymmetry $\beta_c$ for the
disappearance of the mechanical instability is about 0.85 in the case without the
TBF contribution. Inclusion of the TBF reduses $\beta_c$ to about 0.75.
It is also seen that at a fixed asymmetry, the mechanical instable domain of the system
gets smaller as the temperature becomes higher.

In Fig.\ref{fig4} is shown the pressure as a
 function of density at a fixed temperature
$T=5$ MeV for different values of asymmetry $\beta =
0.0, 0.2, 0.4, 0.6, 0.8, 1.0$. In the figure, the solid curves and dashed curves
correspond to the results with and without the TBF contribution.
It can be observed from the figure that the pressure
increases monotonically as the isospin asymmetry increases in agreement with
the results of the RMT calculations\cite{muller:1995} where it is shown that
for asymmetric nuclear matter ($\beta\ne 0$) which is a two-component system,
the pressure cannot remain a constant during the
liquid-gas phase transition.
In the cases of $\beta =0.0$, $0.2$, $0.4$, $0.6$, the system exhibit a
mechanical instability where the pressure is a decreasing function of density.
However for high enough asymmetry, especially for pure neutron matter ($\beta =1$),
the pressure becomes an monotonically increasing function of density,
which indicates the disappearing of the mechanical instability at all
densities. The TBF gives a repulsive contribution in the whole asymmetry range
$0\le \beta\le 1$.

In order to illustrate more clearly the TBF effect
on the properties of the phase transition in asymmetric nuclear matter,
we plot in Fig.\ref{fig5} the domain of mechanical instability
in density-asymmetry plane for the fixed temperature $T=5$ MeV.
In the figure, the solid and dashed curves are obtained by adopting the
$AV_{18}$ plus the TBF and the pure $AV_{18}$ two-body force, respectively.
From the figure we can see that inclusion of the TBF contribution suppresses
considerably the region of mechanical instability in
isothermal asymmetric nuclear matter. In addition, the TBF effect on the
lower-density boundary of the mechanical instable region is very small but
it leads to a remarkable reduction of the upper-density boundary. This is expected
since the TBF repulsion becomes stronger at higher densities.

\section{Summary}

In the present work, we have introduced the microscopic
TBF into the framework of the finite temperature
BHF approach. Employing the extended theoretical model, we
have investigated the EOS and the properties of the liquid-gas phase
transition for hot asymmetric nuclear matter. For symmetric nuclear matter,
the calculated EOS exhibits a clear Van der
Waals behavior in the pressure-density plane, which implies the presence of a
liquid-gas phase transition. The extracted critical temperature is about
16 MeV in the case without the TBF contribution. It is shown that the TBF gives
a repulsive contribution to the EOS and its effect becomes stronger as increasing
density. When the TBF contribution is included, the
critical temperature turns out to be reduced by about 3 MeV from $T_c\approx 16$
MeV to $T_c\approx 13$ MeV as compared to the corresponding two-body force predictions.
The predicted value of $T_c\approx 13$ MeV is close to the value $T_c\simeq 14$ MeV
obtained from the RMT\cite{muller:1995} but it is appreciably larger than that of the relativistic
DB approach.
If including only the $2\sigma-{\rm N}\overline{\rm N}$ contribution of the TBF,
we obtain a lower value of $T_c\approx 11$ MeV which is close to the value
of about 10 MeV predicted by the DB calculations in Ref.\cite{huber:1998}.
 This result is desirable since it may provide a possible explanation for
the discrepancy between the critical temperature values obtained from
 the DB approach and the present BHF plus TBF calculations, i.e, except the
$2\sigma-{\rm N}\overline{\rm N}$ contribution, the other components of the
TBF do not completely cancel among each others at least for the case of
finite temperature and their net effect may becomes more pronounced at higher
temperature. The critical density $\rho_c$ is
found to be roughly $0.065$ fm$^{-3}$ and $0.08$ fm$^{-3}$ in the cases
with and without the TBF contribution, respectively. Both values are
in the range between $\rho_0/3$ and $\rho_0/2$.

For asymmetric nuclear matter, the isospin dependence of the
properties of the liquid-gas phase transition has been studied. It turns out
that the critical temperature decreases and the mechanical instable region
gradually shrinks as the isospin asymmetry $\beta$ increases in both cases
with and without the TBF. The contribution of the TBF to the EOS is repulsive
in the whole range of asymmetry $0\le\beta\le1$ and
results in a reduction of the critical temperature for the phase transition in
hot asymmetric nuclear matter.
Above a critical value of asymmetry, the mechanical
instability disappears for all values of temperature.
The addition repulsion due to the TBF lowers the critical
asymmetry for the disappearance of the mechanical instability
and makes the asymmetric nuclear matter easier to be
gasified. At fixed temperature and density the pressure in
hot asymmetric nuclear system turns out to be an monotonically increasing
function of asymmetry parameter.
In addition, it is also shown that at a fixed asymmetry
the mechanical instable region of hot asymmetric nuclear matter
shrinks as increasing temperature.

\section{Acknowledgment}

One of us (W.Zuo) is very grateful to Prof. T. Gaitanos for helpful
discussions. The work is supported in part by the Chinese Academy of Science Knowledge
Innovation Project (KJCX2-SW-N02), the Major State Basic Research
Development Program of China (G2000077400), and the Major Prophase
Research Project of Fundamental Research of the Ministry of Science and
Technology of China (2002CCB00200), the National Natural Science Foundation
of China (10235030, 10175082) and DFG, Germany.


\newpage

\begin{figure}[tbp]
\caption{Diagrams of the microscopic TBF adopted for the present
calculations, taken from Ref.[26]}
\label{fig1}

\caption{Pressure as a function of density with six isotherms
corresponding to $T=0, 8, 10, 12, 14, 16 MeV$ from the bottom to the top for
different isospin asymmetries $\beta=0, 0.3, 0.5$ and 0.8.
 The solid and dashed curves represent the results
with and without TBF contribution.}
\label{fig2}

\caption{Pressure as a function of density for symmetric nuclear
matter at six values of temperature $T=0, 8, 10, 12, 14$ MeV from
the bottom to the top. The dashed curves are obtained by adopting
the full TBF, while the solid ones by including only the
$2\sigma-{\rm N}\overline{\rm N}$ contribution of the TBF.}
\label{fig3}

\caption{Pressure as a function of density at a fixed temperature $T=5MeV$
for different asymmetric parameters $\beta= 0, 0.2, 0.4, 0.6, 0.8 1.$ from the
bottom to the top. The solid and dashed curves are
the results with and without the TBF contribution, respectively.}
\label{fig4}
\end{figure}

\begin{figure}[tbp]
\caption{Region of mechanical instability at a fixed temperature $T=5MeV$
for both cases including the TBF (solid curve) and without
the TBF (dashed curve).}
\label{fig5}
\end{figure}

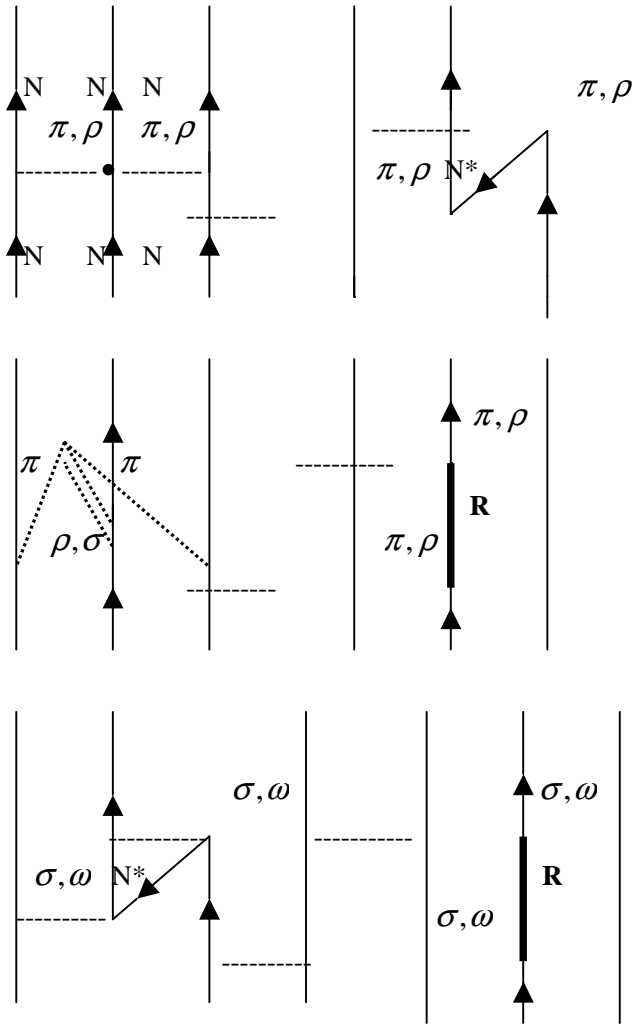
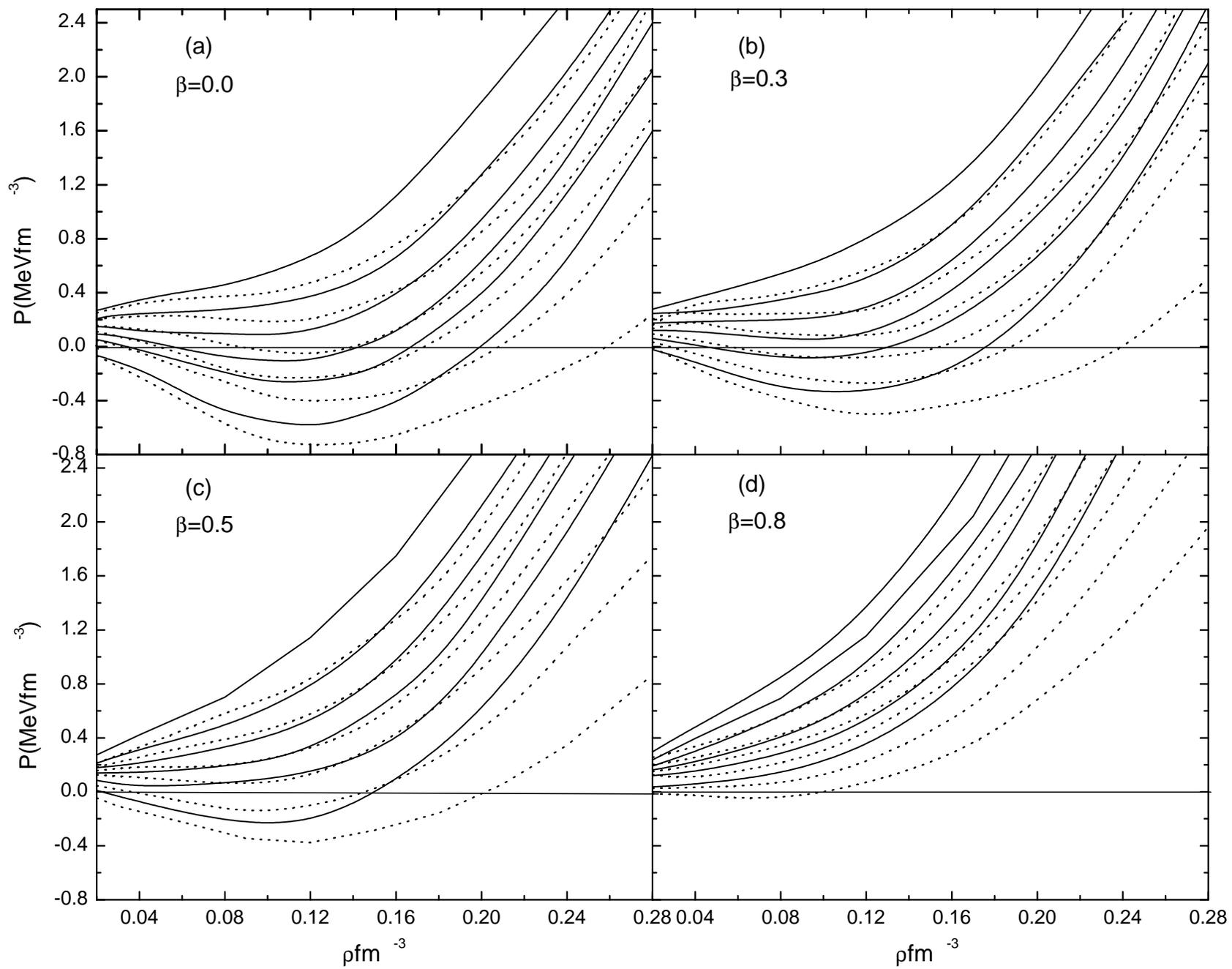
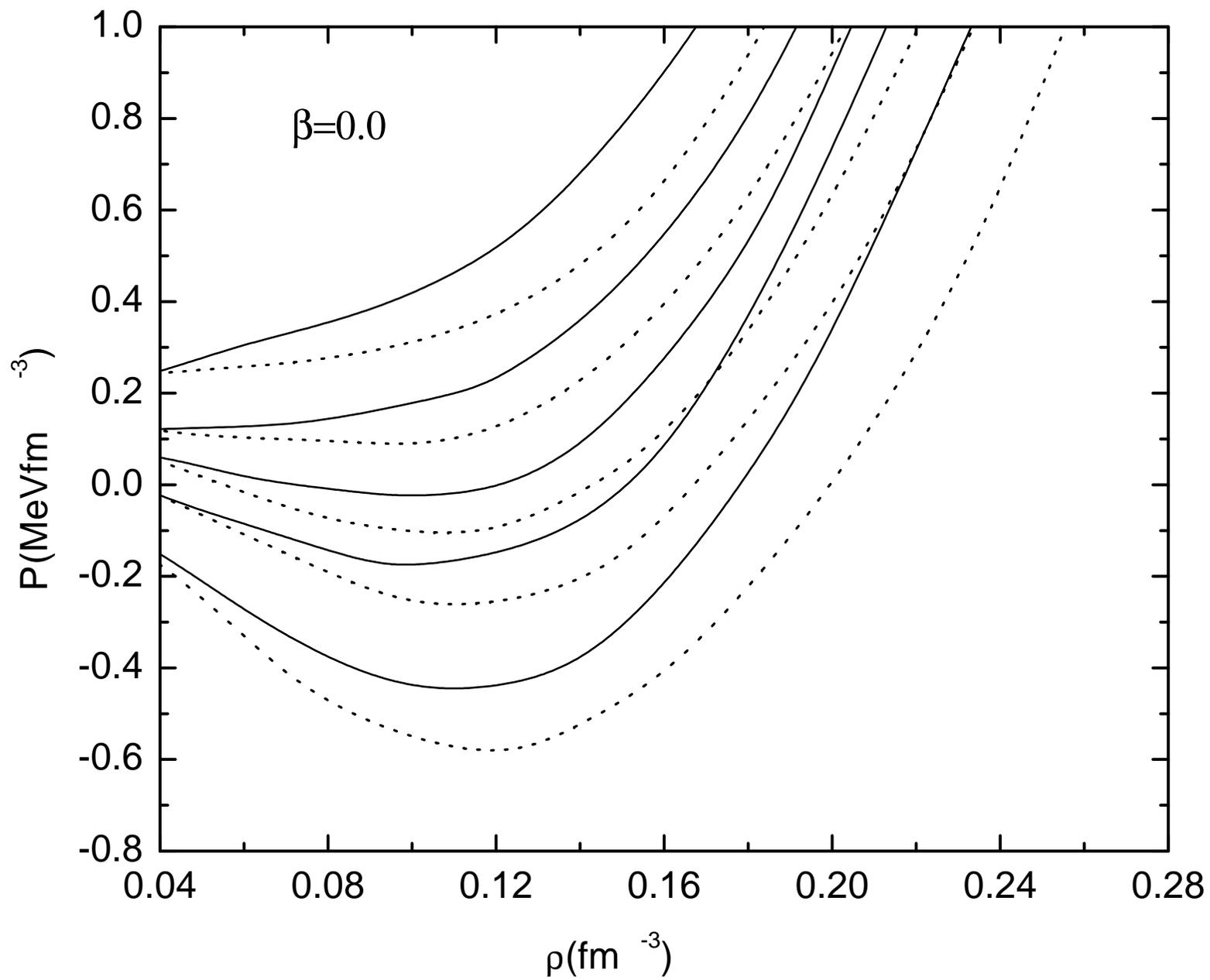
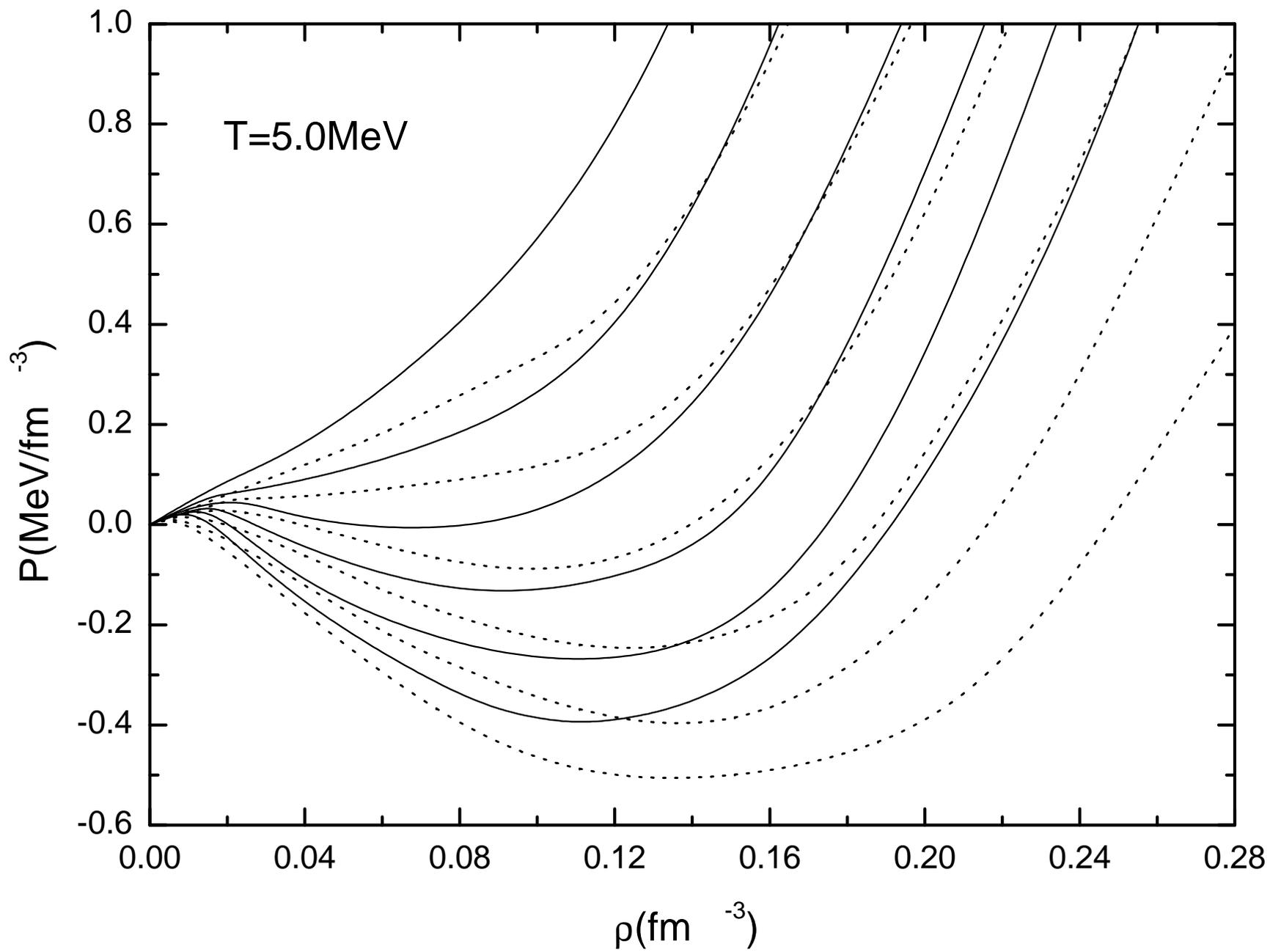
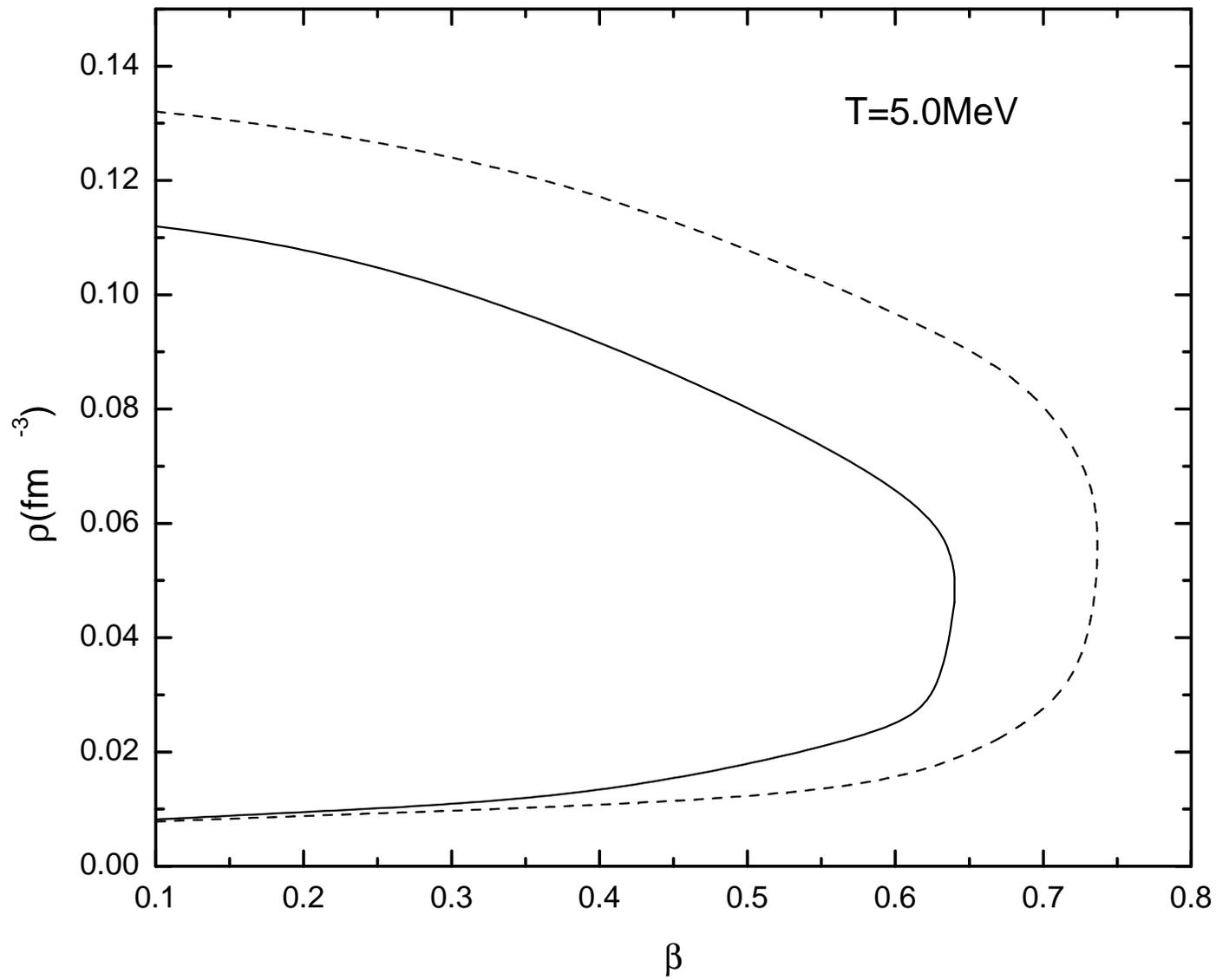
\end{document}